\documentclass[pra]{revtex4-1}

\usepackage{hyperref}
\usepackage[version=3]{mhchem}

\begin{document}

\title{Reply to "Comment on 'Generalization of the Kohn-Sham system that can
       represent arbitrary one-electron density matrices'"}
\author{Hubertus J. J. van Dam}
\affiliation{Brookhaven National Laboratory, Upton, NY 11973-5000}
\date{August 20, 2017}

\begin{abstract}
Our paper [Phys. Rev. A {\bf 93}, 052512 (2016)], proposing a novel form of
single determinant wave function that admits non-idempotent 1-electron 
density matrices, has recently received a Comment [Phys. Rev. A {\bf ??}, 
0????? (2017)] suggesting a number of flaws:
\begin{enumerate}
\item The form of the 1-electron density matrix that we proposed is deemed
      invalid;
\item None of the currently known functionals are given in terms of the
      1-electron density matrix but known only in the basis where the
      density matrix is diagonal;
\item In NOFT the energy is not invariant with respect to unitary 
      transformations of the orbitals;
\item The M{\"u}ller functional we used suffers from serious shortcomings;
\item In NOFT the detachment energies should be obtained from the extended
      Koopmans theorem.
\end{enumerate}
Below we will address these criticisms in sequence. 
\end{abstract}

\maketitle

\section{Introduction}
\label{sect:intro}

In our paper~\cite{vanDam:2016} we proposed a single determinant wave
function and an associated electron probability density so as to enable formulating
an effective 1-electron model that can generate the exact 1-electron density
matrix (1RDM). In addition this wave function requires only $O(N^2)$ coefficients,
where $N$ is the dimension of the basis, so that the memory requirements are
of the same order as those of the 1RDM. Such a 
wave function is desirable for a number of reasons:
\begin{itemize}
\item Single determinant wave functions do not incorporate correlation
      effects and therefore correlation can be accounted for by introducing
      a suitable functional, without risking double counting problems;
\item Single determinant wave functions have easily identified 1-electron
      states facilitating the application of excitation operators;
\item The cost of storing such a wave functions is twice that of 
      storing the Hartree-Fock or Kohn-Sham wave function which is a
      managable overhead;
\item The requirement that the wave function can generate the
      exact electron density, as required by the Hohenberg-Kohn theorems, 
      can be fullfilled whereas with the conventional Kohn-Sham wave function
      this is not strictly possible. 
\end{itemize}
In order to demonstrate the capabilities of this wave function an energy
expression that generates fractionally occupied natural orbitals was needed. 
Rohr~\cite{Rohr:2010} has stated that density functionals usually fail to describe
this kind of correlations well. Instead
the M{\"u}ller functional has been previously demonstrated to favor such 
outcomes. Piris and Pernal, in their comment, explain that there are 
qualitatively better functionals that could
have been used. We accept that such functionals exist and should be considered
for real applications. Nevertheless, for the purpose of demonstrating the
capabilities of the wave function we proposed, the M{\"u}ller functional is
sufficient and its simplicity is beneficial so as not to distract from the main
topic of our publication.

\section{Point 1}

Piris and Pernal [Phys. Rev. A {\bf ??}, 0????? (2017)] claim that my Eq.(16) is
a mistake and invalid, instead they
suggest that their Eq.(5) should be used.
In addition it has been commented that "the 1RDM of a single Slater determinant has
rather specific properties: the orbitals that appear in the Slater determinant are
natural orbitals with occupation one, and the remaining space is described by empty
orbitals." Admittedly, this comment states how we have conventionally used Slater 
determinants, in that the orbitals used are eigenfunctions of the 1RDM and the
associated probability density for finding an electron at position $r$ is given by
\begin{eqnarray}
   p(r) &=& \sum_{a,b=1}^{n_b}\sum_{i=1}^{n_e}\chi_a(r)N_{ai}N^*_{bi}\chi_b^*(r)
\end{eqnarray}
where $n_b$ is the number of basis functions and $n_e$ is the number of electrons 
(considering without loss of generality only spin-up electrons as the properties
for the spin-down electrons can be evaluated in the same way).
However, when Slater introduced the wavefunctions we have come to know as Slater
determinants~\cite{Slater1929} he proposed determinants of spin-orbitals as a convenient
way to formulate wave functions with the proper anti-symmetric behavior under electron
permutations. The key here is that the orbitals are (normalizable) functions of a single
electron (see~\cite{Slater1929} pages 1294, 1299-1300). As orbitals he considered 
1-electron eigen functions of a Hamiltonian with spherical symmetry (page 1299) and
hence it was appropriate to characterize the orbitals by the corresponding quantum
numbers $n$, $l$, $m_l$ and $m_s$. However, as these quantum numbers are specifically 
suited to the study of single atom wavefunctions we may assume that using such orbitals
is not a general requirement.

If our understanding is that a Slater determinant is a wave function represented as
a determinant of 1-electron (spin) wave functions then in principle we have considerable
freedom in the kind of 1-electron wave functions we may choose. The most important
constraint is that the 1-electron wave functions be normalizable. In addition if the 
1-electron wave functions form an orthonormal set we also have that the determinantal
wave function is normalized as long as every orbital appears only once. 

With these considerations in mind we proposed the generalized 1-electron wave function
of our Eq.(8). In addition we deliberately chose our Eq.(16) to express the density matrix
for a single electron. This 1-electron wave function was shown to be normalizable (our
Eqs.(11) to (14)). Hence we maintain that we have satisfied all requirements
for valid 1-electron wave functions.

A determinantal wave function constructed from this (our
Eq.(17)) contains only the occupied generalized orbitals, all other generalized orbitals
remain unoccupied. While the generalized orbitals are either occupied or unoccupied the
correlation functions we introduced distribute a single electron over multiple natural
orbitals. By these means we are able to formulate a single determinantal wave function
that generates a density matrix with fractional occupation numbers. Admittedly these 
choices are rather unconventional but we maintain that unless they lead to inconsistencies
or results that conflict with observations they are permissible.

To provide some background to this choice
consider that in order to change the occupation numbers of a 1RDM (but
without changing the natural orbitals) some additional degrees of freedom are required.
The correlation functions were introduced to provide this additional freedom. Updating
these functions using regular unitary transformations it is ensured that the
orthonormality of the correlation functions is maintained. Now if a Fock matrix of the 
form of our Eq.(9) were used then Eq.(5) of Piris and Pernal would result. In that case,
however, the rotation of correlation functions would just be an additional rotation to
the ones on the natural orbitals. In fact, this would eliminate every possible benefit
of the correlation functions and there would be no reason to introduce them in the first
place, which is main objection of Piris and Pernal.

If instead the Fock
matrix is cast form of our Eq.(44) then the corresponding
density matrix assumes the form of our Eq.(16). Essentially this form entails a 
projection that eliminates off-diagonal terms that would reduce transformations of the
correlation function to simple rotations. Instead the combination of rotation and
projection creates a new kind of transformation. The projection chosen does preserve
critically important properties such as the lower and upper limits on the occupation
numbers as well as the trace of the density matrix. With this choice in place
conventional unitary transformations ensure
that orthonormality of the 1-electron wave functions is preserved, the trace and
eigenvalue limits
of the density matrix are preserved, yet non-trivial transformations that are cannot
be formaluted in an N-dimensional space can now be expressed straightforwardly on the
1RDM. A worked example of this approach is discussed in the 
supplementary material, Section I, where it is shown how this technique can used to 
transform a 1RDM of a single electron from
\begin{equation}
   D = \left(\begin{array}{cc}
             1 & 0 \\
             0 & 0 
             \end{array}\right)
\end{equation}
to 
\begin{equation}
   D' = \left(\begin{array}{cc}
              1/2 & 0 \\
              0   & 1/2
              \end{array}\right)
\end{equation}
It is also shown that such a transformation cannot be represented in a 2-dimensional
space.

Piris and Pernal have also raised the issue of the N-representability of NOFT and 
pointed to the fact that attempts to express the 2RDM by means of a reconstruction 
functional have failed (we will come back to that in Section~\ref{point2}). 
The reason being that the ensemble N-representability 
conditions of the 1RDM are easily implemented but that they are not sufficient to 
guarantee that the reconstructed 2RDM is N-representable. As we believe that we can
maintain an approach until properly falsified and as a wave function should by
construction generate N-representable density matrices let us consider the same spin 
part of the 2RDM as a test for our proposed approach (the opposite spin part is 
essentially trivial as the exchange terms vanish). 

To summarize a single determinant wave function is defined of the form (Eq.(17)
of~\cite{vanDam:2016}):
\begin{equation}
   \Psi(r_1,\ldots,r_{n_e})
   = \left|G_1(r_1)G_2(r_2)\ldots G_{n_e}(r_{n_e})\right|
   \label{Eq:wavefunction}
\end{equation}
where $G(r)$ are the generalized orbitals we proposed given by (Eq.(8)
of~\cite{vanDam:2016}):
\begin{eqnarray}
   \left|G_s(r)\right\rangle
   &=& \sum_{a,i=1}^{n_b} N_{ai}C_{is}\left|\chi_a(r)\right\rangle
   \label{Eq:genorb}
\end{eqnarray}
The 1-electron density matrix of Eq.(16) of~\cite{vanDam:2016} is adopted as given by
\begin{eqnarray}
   \sum_{a,b=1}^{n_b}\left|\chi_a(r'_1)\right\rangle D_{ab}
                     \left\langle\chi_b(r''_1)\right|
   &=& \left(\begin{array}{c}n_e \\ 1\end{array}\right)
       \int\ldots\int \Psi(r'_1,r_2,\ldots,r_{n_e})\Psi^*(r''_1,r_2,\ldots,r_{n_e})
       \mathrm{d}r_2 \ldots \mathrm{d}r_{n_e}
       \label{Eq:1RDM_1} \\
   D_{ab} 
   &=& \sum_{i=1}^{n_b}\sum_{s=1}^{n_e} N_{ai}C_{is}C^*_{is}N^*_{bi}
       \label{Eq:1RDM}
\end{eqnarray}
furthermore the 2RDM is defined by
\begin{eqnarray}
   &\sum_{a,b,c,d=1}^{n_b}&\left|\chi_a(r'_1)\chi_b(r'_2)\right\rangle
                         \Gamma_{abcd}
                         \left\langle\chi_c(r''_1)\chi_d(r''_2)\right|
                         \nonumber \\
   &&= \left(\begin{array}{c}n_e \\ 2\end{array}\right)
       \int\ldots\int
       \Psi(r'_1,r'_2,\ldots,r_{n_e})\Psi^*(r''_1,r''_2,\ldots,r_{n_e})
       \mathrm{d}r_3 \ldots \mathrm{d}r_{n_e}
       \label{Eq:2RDM_1} 
\end{eqnarray}
Obviously, Eq.(\ref{Eq:1RDM}) cannot be used directly in evaluating $\Gamma_{abcd}$ as
for the 2RDM we have to consider cross terms between orbitals. To accomodate these 
cross terms Eq.(\ref{Eq:1RDM}) is trivially generalized as
\begin{eqnarray}
   D^{st}_{ab} 
   &=& \sum_{i=1}^{n_b} N_{ai}C_{is}C^*_{it}N^*_{bi}
       \label{Eq:1RDMp}
\end{eqnarray}
Note that because of the orthonormality of the correlation functions the trace of 
Eq.(\ref{Eq:1RDMp}) is
\begin{eqnarray}
   \mathrm{tr}\left(D^{st}\right) &=& \delta_{st}
   \label{Eq:trace}
\end{eqnarray}
which will play an important role in considering the exchange terms.
In addition as the generalized orbitals are defined as in Eq.(\ref{Eq:genorb})
anti-symmetric permutations among generalized orbitals have to be defined as
\begin{eqnarray}
   P(1,2)G_1(r_1)G_2(r_2)
   &=& P(1,2)\sum_{a,b=1}^{n_b}\sum_{i,j=1}^{n_b}N_{ai}
       C_{i1}\chi_a(r_1)N_{bj}C_{j2}\chi_b(r_2) \\
   &=& -\sum_{a,b=1}^{n_b}\sum_{i,j=1}^{n_b}N_{ai}
        C_{i1}\chi_a(r_2)N_{bj}C_{j2}\chi_b(r_1) 
        \label{Eq:permute} \\
\end{eqnarray}
in other words the permutation operator implies exchanging units of $\sum_{i}N_{ai}C_{is}$
in its entirety. Alternatives such as exchanging only the correlation functions are not 
valid as they do not treat the entire generalized orbital as an integral unit. Based on
Eq.(\ref{Eq:2RDM_1}), Eq.(\ref{Eq:1RDMp}) and Eq.(\ref{Eq:permute}) the 2RDM can be
written as
\begin{eqnarray}
   \Gamma_{abcd} 
   &=& \frac{1}{4}\left\{
      \sum_{s,t=1}^{n_e}
      \left(\sum_{i=1}^{n_b} N_{ai}C_{is}C^*_{is}N^*_{ci}\right)
      \left(\sum_{j=1}^{n_b} N_{bj}C_{jt}C^*_{jt}N^*_{dj}\right)
      \right.
      \nonumber \\
  &&- \sum_{s,t=1}^{n_e}
      \left(\sum_{i=1}^{n_b} N_{ai}C_{is}C^*_{it}N^*_{di}\right)
      \left(\sum_{j=1}^{n_b} N_{bj}C_{jt}C^*_{js}N^*_{cj}\right)
      \nonumber \\
  &&- \sum_{s,t=1}^{n_e}
      \left(\sum_{i=1}^{n_b} N_{bi}C_{it}C^*_{is}N^*_{ci}\right)
      \left(\sum_{j=1}^{n_b} N_{aj}C_{js}C^*_{jt}N^*_{dj}\right)
      \nonumber \\
  &&+ \left.
      \sum_{s,t=1}^{n_e}
      \left(\sum_{i=1}^{n_b} N_{bi}C_{it}C^*_{it}N^*_{di}\right)
      \left(\sum_{j=1}^{n_b} N_{aj}C_{js}C^*_{js}N^*_{cj}\right)
      \right\}
      \label{Eq:2RDM}
\end{eqnarray}
Given the expressions above the N-representability conditions may be addressed.
They are:
\begin{enumerate}
\item The trace of the 2RDM must be equal to the number of electron pairs
      $\frac{n_e(n_e-1)}{2}$;
\item The matrix must non-negative, i.e. all eigenvalues must be equal to or larger
      than $0$, or equivalently all electron pair probability functions must be
      non-negative;
\item Integrating the coordinates of 1 electron out, must produce the 1RDM according to
      $D_1(r_1;r'_1) = \frac{2}{N-1}\int D_2(r_1,r_2;r'_1,r_2)\mathrm{d}r_2$
\item The 2RDM must be anti-symmetric, i.e. 
      $D_2(r_1,r_2;r'_1,r'_2)=-D_2(r_1,r_2;r'_2,r'_1)=-D_2(r_2,r_1;r'_1,r'_2)=D_2(r_2,r_1;r'_2,r'_1)$
\end{enumerate}
Below this points are tackeled in turn.

\subsection{The trace of the 2RDM}

The trace of the 2RDM given by Eq.(\ref{Eq:2RDM}) can be evaluated by considering and 
adding the traces of the various terms because
$\mathrm{tr}(A+B)=\mathrm{tr}(A)+\mathrm{tr}(B)$ based simply on the fact that the 
trace of a matrix is the sum of its diagonal elements. For the first and the fourth term
of Eq.(\ref{Eq:2RDM}) the traces are given by 
\begin{eqnarray}
   \mathrm{tr}(\mathrm{term_1})
   &=& \sum_{s,t=1}^{n_e}\sum_{i,j=1}^{n_b} C_{is}C^*_{is}C_{jt}C^*_{jt} \\
   &=& \sum_{s,t=1}^{n_e}\delta_{ss}\delta_{tt} \\
   &=& n_e^2 \label{Eq:1term1} \\
   \mathrm{tr}(\mathrm{term_4})
   &=& \sum_{s,t=1}^{n_e}\sum_{i,j=1}^{n_b} C_{it}C^*_{it}C_{js}C^*_{js} \\
   &=& n_e^2 \label{Eq:1term4}
\end{eqnarray}
For the second and third terms of Eq.(\ref{Eq:2RDM}) the traces
\begin{eqnarray}
   \mathrm{tr}(\mathrm{term_2})
   &=& -\sum_{s,t=1}^{n_e}\sum_{i,j=1}^{n_b} C_{is}C^*_{it}C_{jt}C^*_{js} \\
   &=& -\sum_{s,t=1}^{n_e}\delta_{st}\delta_{ts} \\
   &=& -n_e \label{Eq:1term2} \\
   \mathrm{tr}(\mathrm{term_3})
   &=& -\sum_{s,t=1}^{n_e}\sum_{i,j=1}^{n_b} C_{it}C^*_{is}C_{js}C^*_{jt} \\
   &=& -n_e \label{Eq:1term3} \\
\end{eqnarray}
are obtained.
Combining the results of Eqs.(\ref{Eq:1term1}, \ref{Eq:1term4}, \ref{Eq:1term2},
\ref{Eq:1term3}) with the prefactor of Eq.(\ref{Eq:2RDM}) the result
\begin{eqnarray}
   \mathrm{tr}\left(\Gamma_{abcd}\right)
   &=& \frac{1}{4}\left(n_e^2-n_e-n_e+n_e^2\right) \\
   &=& \frac{n_e(n_e-1)}{2} \label{Eq:proof1}
\end{eqnarray}
is obtained which proves that the first N-presentability condition is met.

\subsection{The non-negativity of the 2RDM}

To prove that the 2RDM of Eq.(\ref{Eq:2RDM}) is non-negative it has to be shown
that the Coulomb terms (terms 1 and 4) are larger than the exchange terms
(terms 2 and 3). The key is to show that the diagonal elements are non-negative. Once
that is established one can always show that the pair probability density is 
non-negative as well, which implies that the density matrix must be non-negative. 
The proof is based on the Gauchy-Schwarz inequality which is given by
\begin{eqnarray}
   \left(\sum_{i=1}^n x_i y_i^*\right)^2 
   &\leq& \left(\sum_{j=1}^n x_j^2\right)\left(\sum_{k=1}^n y_k^2\right)
          \label{Eq:GauchySchwarz}
\end{eqnarray}
The diagonal elements of Eq.(\ref{Eq:2RDM}) are given by
\begin{eqnarray}
   d_{ij} &=& \frac{1}{4}\sum_{s,t=1}^{n_e}\left(
              C_{is}C^*_{is}C_{jt}C^*_{jt}-C_{is}C^*_{it}C_{jt}C^*_{js}
             -C_{it}C^*_{is}C_{js}C^*_{jt}+C_{it}C^*_{it}C_{js}C^*_{js}\right)
\end{eqnarray}
Obviously terms 1 and 4 and terms 2 and 3 are essentially the same and therefore the
expression can be reduced to two terms as
\begin{eqnarray}
   d_{ij} &=& \frac{1}{2}\sum_{s,t=1}^{n_e}\left(
              C_{is}C^*_{is}C_{jt}C^*_{jt}-C_{is}C^*_{it}C_{jt}C^*_{js}\right)
\end{eqnarray}
Rewriting this expression obtains
\begin{eqnarray}
   d_{ij} &=& \frac{1}{2}\left[
              \left(\sum_{s=1}^{n_e}C_{is}C^*_{is}\right)
              \left(\sum_{t=1}^{n_e}C_{jt}C^*_{jt}\right)
             -\left(\sum_{s=1}^{n_e}C_{is}C^*_{js}\right)
              \left(\sum_{t=1}^{n_e}C_{jt}C^*_{it}\right)\right] \\
          &=& \frac{1}{2}\left[
              \left(\sum_{s=1}^{n_e}C_{is}C^*_{is}\right)
              \left(\sum_{t=1}^{n_e}C_{jt}C^*_{jt}\right)
             -\left(\sum_{s=1}^{n_e}C_{is}C^*_{js}\right)^2\right]
\end{eqnarray}
From the Gauchy-Schwarz inequality one immediately obtains that 
\begin{eqnarray}
   d_{ij} \geq 0 \label{Eq:GSdiag}
\end{eqnarray}
Obviously the argument given leaves the factors stemming from the natural orbitals out
and as they are different in different terms this makes the outcome less clear.
To resolve this matter the pair probability density is considered
\begin{eqnarray}
   p(r_1,r_2)
   &=& \frac{1}{2}\sum_{s,t=1}^{n_e}\left[
       \left(\sum_{a,c=1}^{n_b}\sum_{i=1}^{n_b}
             \chi_a(r_1)N_{ai}C_{is}C^*_{is}N^*_{ci}\chi^*_c(r_1)
       \right)
       \left(\sum_{b,d=1}^{n_b}\sum_{j=1}^{n_b}
             \chi_b(r_2)N_{bj}C_{jt}C^*_{jt}N^*_{dj}\chi^*_d(r_2)
       \right)\right. \nonumber \\
   &&- \left.\left(\sum_{a,d=1}^{n_b}\sum_{i=1}^{n_b}
             \chi_a(r_1)N_{ai}C_{is}C^*_{it}N^*_{di}\chi^*_d(r_1)
       \right)
       \left(\sum_{b,c=1}^{n_b}\sum_{j=1}^{n_b}
             \chi_b(r_2)N_{bj}C_{jt}C^*_{js}N^*_{cj}\chi^*_c(r_2)
       \right)
       \right]
\end{eqnarray}
Relabeling the basis function indeces in the exchange terms gives
\begin{eqnarray}
   p(r_1,r_2)
   &=& \frac{1}{2}\sum_{s,t=1}^{n_e}\left[
       \left(\sum_{a,c=1}^{n_b}\sum_{i=1}^{n_b}
             \chi_a(r_1)N_{ai}C_{is}C^*_{is}N^*_{ci}\chi^*_c(r_1)
       \right)
       \left(\sum_{b,d=1}^{n_b}\sum_{j=1}^{n_b}
             \chi_b(r_2)N_{bj}C_{jt}C^*_{jt}N^*_{dj}\chi^*_d(r_2)
       \right)\right. \nonumber \\
   &&- \left.\left(\sum_{a,c=1}^{n_b}\sum_{i=1}^{n_b}
             \chi_a(r_1)N_{ai}C_{is}C^*_{it}N^*_{ci}\chi^*_c(r_1)
       \right)
       \left(\sum_{b,d=1}^{n_b}\sum_{j=1}^{n_b}
             \chi_b(r_2)N_{bj}C_{jt}C^*_{js}N^*_{dj}\chi^*_d(r_2)
       \right)
       \right]
       \label{Eq:2probability}
\end{eqnarray}
At which point Eq.(\ref{Eq:GSdiag}) can be used to demonstrate that
\begin{eqnarray}
   p(r_1,r_2) \geq 0 \label{Eq:2prob_a}
\end{eqnarray}
For the Gauchy-Schwarz inequality to hold there is no need to integrate over all 
electrons. The inequality also holds for every term separately, i.e. $n=1$ in 
Eq.(\ref{Eq:GauchySchwarz}). Hence we also have
\begin{eqnarray}
   p^{st}(r_1,r_2)
   &=& \frac{1}{2}\left[
       \left(\sum_{a,c=1}^{n_b}\sum_{i=1}^{n_b}
             \chi_a(r_1)N_{ai}C_{is}C^*_{is}N^*_{ci}\chi^*_c(r_1)
       \right)
       \left(\sum_{b,d=1}^{n_b}\sum_{j=1}^{n_b}
             \chi_b(r_2)N_{bj}C_{jt}C^*_{jt}N^*_{dj}\chi^*_d(r_2)
       \right)\right. \nonumber \\
   &&- \left.\left(\sum_{a,c=1}^{n_b}\sum_{i=1}^{n_b}
             \chi_a(r_1)N_{ai}C_{is}C^*_{it}N^*_{ci}\chi^*_c(r_1)
       \right)
       \left(\sum_{b,d=1}^{n_b}\sum_{j=1}^{n_b}
             \chi_b(r_2)N_{bj}C_{jt}C^*_{js}N^*_{dj}\chi^*_d(r_2)
       \right)
       \right]
       \label{Eq:2probability_st}
\end{eqnarray}
At which point Eq.(\ref{Eq:GSdiag}) can be used to demonstrate all electron pair
probability functions are non-negative
\begin{eqnarray}
   p^{st}(r_1,r_2) \geq 0, \;  \forall s,t \le n_e \label{Eq:proof2}
\end{eqnarray}

The result of Eq.(\ref{Eq:proof2}) proves that the second N-representability 
condition is met.

\subsection{Obtaining the 1RDM by integration of the 2RDM}

The integration of the one set of coordinates from the 2RDM of Eq.(\ref{Eq:2RDM}) to
calculate the 1RDM can be tackled term-by-term as well. Note that here only the 
evaluation of the $\alpha$-electron 1RDM from the $\alpha\alpha$-part of the 2RDM
is considered. This differs by a constant from the calculation of the $\alpha$-1RDM from 
the whole 2RDM. In the latter case there is a term coming from the $\alpha\alpha$ block
as well as from the $\alpha\beta$ block. Calculating the $\alpha$-1RDM from the 
$\alpha\beta$ block is trivial hence the more involved calculation from the 
$\alpha\alpha$ block is considered here. Use is made of the fact
that the natural orbitals are normalized such that
\begin{eqnarray}
   \delta_{ij} &=& \sum_{a,b=1}^{n_b} N_{ai}\langle\chi_a(r_1)|\chi_b(r_1)\rangle N_{bj}
\end{eqnarray}
Integrating out electron 2 for
the first and fourth term gives
\begin{eqnarray}
   T_1
   &=& \left(\sum_{i=1}^{n_b}\sum_{s=1}^{n_e}N_{ai}C_{is}C^*_{is}N_{ci}\right)
       \left(\sum_{j=1}^{n_b}\sum_{t_1}^{n_e}\sum_{b,d=1}^{n_b}
             N_{bj}C_{jt}C^*_{jt}N^*_{dj}\langle\chi_b(r_2)|\chi_d(r_2)\rangle\right) \\
   &=& \left(\sum_{i=1}^{n_b}\sum_{s=1}^{n_e}N_{ai}C_{is}C^*_{is}N_{ci}\right)
       \left(\sum_{j=1}^{n_b}\sum_{t_1}^{n_e}
             C_{jt}C^*_{jt}\right) \\
   &=& \left(\sum_{i=1}^{n_b}\sum_{s=1}^{n_e}N_{ai}C_{is}C^*_{is}N_{ci}\right)n_e \\
   &=& D_{ac}n_e \label{Eq:3term1} \\
   T_4
   &=& \left(\sum_{i=1}^{n_b}\sum_{t=1}^{n_e}N_{bi}C_{it}C^*_{it}N_{di}\right)
       \left(\sum_{j=1}^{n_b}\sum_{s_1}^{n_e}\sum_{a,c=1}^{n_b}
             N_{aj}C_{js}C^*_{js}N^*_{cj}\langle\chi_a(r_2)|\chi_c(r_2)\rangle\right) \\
   &=& \left(\sum_{i=1}^{n_b}\sum_{t=1}^{n_e}N_{bi}C_{it}C^*_{it}N_{di}\right)
       \left(\sum_{j=1}^{n_b}\sum_{s_1}^{n_e}
             C_{js}C^*_{js}\right) \\
   &=& \left(\sum_{i=1}^{n_b}\sum_{t=1}^{n_e}N_{bi}C_{it}C^*_{it}N_{di}\right)n_e \\
   &=& D_{bd}n_e \label{Eq:3term4}
\end{eqnarray}
Integrating out electron 2 for the second and third term gives
\begin{eqnarray}
   T_2
   &=& - \sum_{s,t=1}^{n_e}
         \left(\sum_{i=1}^{n_b} N_{ai}C_{is}C^*_{it}N^*_{di}\right)
         \left(\sum_{j=1}^{n_b} \sum_{b,c=1}^{n_b} 
               N_{bj}C_{jt}C^*_{js}N^*_{cj}
               \langle\chi_b(r_2)|\chi_c(r_2)\rangle\right) \\
   &=& - \sum_{s,t=1}^{n_e}
         \left(\sum_{i=1}^{n_b} N_{ai}C_{is}C^*_{it}N^*_{di}\right)
         \left(\sum_{j=1}^{n_b} C_{jt}C^*_{js} \right) \\
   &=& - \sum_{s,t=1}^{n_e}
         \left(\sum_{i=1}^{n_b} N_{ai}C_{is}C^*_{it}N^*_{di}\right)\delta_{ts} \\
   &=& - D_{ad} \label{Eq:3term2} \\
   T_3
   &=& - \sum_{s,t=1}^{n_e}
         \left(\sum_{i=1}^{n_b} N_{bi}C_{it}C^*_{is}N^*_{ci}\right)
         \left(\sum_{j=1}^{n_b}\sum_{a,d=1}^{n_b}
               N_{aj}C_{js}C^*_{jt}N^*_{dj}
               \langle\chi_a(r_2)|\chi_d(r_2)\rangle\right) \\
   &=& - \sum_{s,t=1}^{n_e}
         \left(\sum_{i=1}^{n_b} N_{bi}C_{it}C^*_{is}N^*_{ci}\right)
         \left(\sum_{j=1}^{n_b}\sum_{a,d=1}^{n_b} C_{js}C^*_{jt} \right) \\
   &=& - \sum_{s,t=1}^{n_e}
         \left(\sum_{i=1}^{n_b} N_{bi}C_{it}C^*_{is}N^*_{ci}\right)\delta_{st} \\
   &=& - D_{bc} \label{Eq:3term3}
\end{eqnarray}
Combining the results of the Eqs.(\ref{Eq:3term1}, \ref{Eq:3term4}, \ref{Eq:3term2},
\ref{Eq:3term3}) with the prefactor gives
\begin{eqnarray}
   D
   &=& \frac{1}{4}\frac{2}{n_e-1}\left(n_e D - D - D + n_e D\right) \\
   &=& \frac{1}{4}\frac{2}{n_e-1}2 D (n_e-1) \\
   &=& D \label{Eq:proof3}
\end{eqnarray}
This result proves that the third N-representability condition is met.

\subsection{The anti-symmetry of the 2RDM}

In order to demonstrate the anti-symmetry of the 2RDM the condition given in 
Eq.(\ref{Eq:permute}) for the permutation symmetry is enforced. This means that
exchanging two labels requires exchanging the whole orbital expression. Note that
this is the same condition we enforced in the 2RDM construction.
With that
\begin{eqnarray}
   \Gamma_{abdc} 
   &=& \frac{1}{4}\left\{
      \sum_{s,t=1}^{n_e}
      \left(\sum_{i=1}^{n_b} N_{ai}C_{is}C^*_{it}N^*_{di}\right)
      \left(\sum_{j=1}^{n_b} N_{bj}C_{jt}C^*_{js}N^*_{cj}\right)
      \right.
      \nonumber \\
  &&- \sum_{s,t=1}^{n_e}
      \left(\sum_{i=1}^{n_b} N_{ai}C_{is}C^*_{is}N^*_{ci}\right)
      \left(\sum_{j=1}^{n_b} N_{bj}C_{jt}C^*_{jt}N^*_{dj}\right)
      \nonumber \\
  &&- \sum_{s,t=1}^{n_e}
      \left(\sum_{i=1}^{n_b} N_{bi}C_{it}C^*_{it}N^*_{di}\right)
      \left(\sum_{j=1}^{n_b} N_{aj}C_{js}C^*_{js}N^*_{cj}\right)
      \nonumber \\
  &&+ \left.
      \sum_{s,t=1}^{n_e}
      \left(\sum_{i=1}^{n_b} N_{bi}C_{it}C^*_{is}N^*_{ci}\right)
      \left(\sum_{j=1}^{n_b} N_{aj}C_{js}C^*_{jt}N^*_{dj}\right)
      \right\} \\
  &=& -\frac{1}{4}\left\{
      \sum_{s,t=1}^{n_e}
      \left(\sum_{i=1}^{n_b} N_{ai}C_{is}C^*_{is}N^*_{ci}\right)
      \left(\sum_{j=1}^{n_b} N_{bj}C_{jt}C^*_{jt}N^*_{dj}\right)
      \right.
      \nonumber \\
  &&- \sum_{s,t=1}^{n_e}
      \left(\sum_{i=1}^{n_b} N_{ai}C_{is}C^*_{it}N^*_{di}\right)
      \left(\sum_{j=1}^{n_b} N_{bj}C_{jt}C^*_{js}N^*_{cj}\right)
      \nonumber \\
  &&- \sum_{s,t=1}^{n_e}
      \left(\sum_{i=1}^{n_b} N_{bi}C_{it}C^*_{is}N^*_{ci}\right)
      \left(\sum_{j=1}^{n_b} N_{aj}C_{js}C^*_{jt}N^*_{dj}\right)
      \nonumber \\
  &&+ \left.
      \sum_{s,t=1}^{n_e}
      \left(\sum_{i=1}^{n_b} N_{bi}C_{it}C^*_{it}N^*_{di}\right)
      \left(\sum_{j=1}^{n_b} N_{aj}C_{js}C^*_{js}N^*_{cj}\right)
      \right\} \\
  &=& -\Gamma_{abcd} \label{Eq:proof4}
\end{eqnarray}
With Eq.(\ref{Eq:proof4}) it is demonstrated that the fourth N-representability 
condition is met. 

\subsection{Summary of 2RDM results}

Based on the arguments above it would seem that the approach proposed
in~\cite{vanDam:2016} provides both N-representable 1RDM and 2RDM without generating
logical inconsistencies. Therefore even applying our approach to the 2RDM does not
lead to a breakdown that invalidates the wave function we proposed.
In addition to
ref.~\cite{vanDam:2016} the exact 1RDM of a system of interacting electrons can be 
represented. Of course the exact 2RDM of a system of interacting electrons cannot be
obtained as such an RDM would have to account for electron correlation (the approach
suggested here is merely a Hartree-Fock theory with funky orbitals, i.e. one that admits
fractional occupations of the natural orbitals). As the wave function used
here is just a single determinant wave function it cannot capture electron correlation
at all. This is also clear from the fact that the 2RDM presented here has 
anti-symmetrized products of
natural orbitals as eigenfunctions rather than proper geminals. In addition the pair
occupation numbers are limited to the range $[0,1]$ whereas in the exact 2RDM they
may exceed $1$.~\cite{Sasaki1965}

In fact if the wave function we proposed in~\cite{vanDam:2016} is simply inserted
into the Schr{\"o}dinger equation and the energy minimized a wave function that is
equivalent to the conventional Hartree-Fock wave function will be obtained.
To account for electron correlation special terms 
that model electron correlation as a function of the natural orbital occupation numbers
would have to be introduced. In NOFT and~\cite{vanDam:2016} the 2-electron interaction
is effectively replaced with a functional of the natural orbitals and occupation numbers.

\section{Point 2}
\label{point2}

To address the criticism that none of the currently known functionals are
given in terms of the 1RDM but known only in the basis
where the density matrix is diagonal two things need to be considered:
\begin{itemize}
\item The equivalence between functions expressed on eigenvalues and matrix
      functions
\item Show that multiple 1RMD functionals have been
      proposed that can be recast in such a way as to rely on matrix functions
\end{itemize}

The first point is a well know result from the mathematics of matrix functions. 
If $M$ is a Hermitian matrix and $\{v_1,...,v_i,...,v_n\}$ are its eigenvectors
with eigenvalues $\{d_1,...,d_i,...,d_n\}$ then a function $f$ applied to the
matrix can be obtained as 
\begin{equation}
   f(M) = \sum_{i=1}^{n} v_i f(d_i) v^T_i
   \label{Eq:FM}
\end{equation}
This is a well known property of matrix functions~\cite{Higham:2008}. From this
we have that even if a 1-electron density matrix functional is given in terms
of functions applied to occupation numbers then, as long as the same function
is applied to all occupation numbers, this function can be rewritten as 
a function on the 1RDM. In the latter form no
explicit mention of the occupation numbers is even needed.

For the second point we need show that there are multiple 1RDM functionals
that are expressed either directly in terms of the 1RDM
or (equivalently) that are expressed in terms of functions on the occupation
numbers where the same function is applied to all occupation numbers. Some 
examples of such functionals are:
\begin{itemize}
\item First of all Piris and Pernal have admitted that the Hartree-Fock energy
      expression is an explicit density matrix functional, albeit a rather
      trivial one and one that does not account for electron correlation at all.
\item The M{\"u}ller functional~\cite{Muller1984} and variants 
      thereof~\cite{Lathiotakis2009}.
\item The functional of Cs{\'a}nyi and Arias~\cite{Csanyi:2000} where the
      authors explicitly state on page 7350 in the first paragraph of the
      subsection ``Representation of the functionals'' that ``The corresponding
      energy functionals may be represented either directly in terms of the
      one-body density matrix for use with direct density-matrix methods, or in
      terms of the spectral (natural orbital representation of the density
      matrix.''
\item Herbert and Harriman~\cite{Herbert:2003} consider a range of different
      functionals including direct and implicit density matrix functionals
      (including the ones by M{\"u}ller and Cs{\'a}nyi and generalizations
      thereof). They refer to these functionals as $CH(\zeta)$,$CHF(\zeta)$,
      and $MCHF(\zeta)$.
\end{itemize}
Admittedly this list is not a very extensive one 
but at least there are a few. Hence the claim that no such
functionals exist seems exaggerated.

\section{Point 3}

The criticism that the energy in NOFT is not invariant to orbital
transformations seems to need some additional information. The authors 
Piris and Pernal make distinctions between implicit and explicit 1-electron
density functionals. Based on what we have seen regarding this issue we 
conclude that Piris and Pernal refer to explicit 1-electron density matrix
functionals when a functional can be represented in a form that involves 
functions of the 1RDM as in the left-hand-side of Eq.(\ref{Eq:FM}). In functionals
where different functions may be applied to different occupation numbers, 
i.e. where the functional cannot be expressed using matrix functions, they
refer to those expressions as implicit 1RMD functionals or Natural Orbital Functionals
(NOF).

We have no experience or expertise related to implicit 1RDM
functionals and therefore we will not comment on those functionals and
any specific requirements related to their energy optimization. On the topic
of explicit 1RDM functionals Piris and Pernal agree that the optimization
method we used is valid. As for the L{\"o}wdin paper~\cite{Lowdin:1955} 
%
%
this problem applies to the
Langrangian associated with the natural orbitals but not to the Lagrangian
of the correlation functions. The correlation functions are either occupied
or unoccupied. The 1RDM is invariant under
rotations amongst the occupied correlation functions, and invariant under
rotations amongst the unoccupied correlation functions. This leaves ample 
freedom to diagonalize the matrix of Lagrangian multipliers for the 
correlation functions. In addition the correlation function energies (or
equivalently the generalized orbital energies) are the proper one-electron
energies. Hence the duality issue that Piris and Pernal face in their 
natural orbital driven formulation does not arise here.

\section{Point 4}

As stated in Section~\ref{sect:intro} we accept that the M{\"u}ller 
functional has short comings. It was only chosen for its simplicity 
to demonstrate the
ability to generate fractional occupation numbers with the approach
proposed in our paper~\cite{vanDam:2016}. For real applications 
other functionals should be chosen. We do not claim any expertise
on the question which functional is most appropriate in 
reduced density matrix methods.

\section{Point 5}

The issue with the detachment energies is that the orbital energies
of a correlated electron system are the same for all correlated electrons.
In a way this suggests a collective behavior of the correlated electrons and
therefore the associated energy is not a good measure for a single electron
response. Hence we proposed to calculate these single electron energies
from an effective 1-electron (i.e. uncorrelated) energy expression instead.
This choice does not preclude that there other and possibly better ways 
to do this.

\begin{acknowledgments}
This manuscript has been authored by employees of Brookhaven Science Associates,
LLC under Contract No. DE- SC0012704 with the U.S. Department of Energy.
\end{acknowledgments}

\bibliographystyle{unsrt}
\bibliography{REPLY_TO}

\end{document}